\documentstyle[multicol,aps,psfig]{revtex}
\renewcommand{\narrowtext}{\begin{multicols}{2}
\global\columnwidth20.5pc\noindent}
\renewcommand{\widetext}{\end{multicols}
\global\columnwidth42.5pc}
\begin{document}
\draft
\preprint{4 March 2004}
\title{Nuclear Spin-Lattice Relaxation in One-Dimensional
       Heisenberg Ferrimagnets:\\
       Three-Magnon versus Raman Processes}
\author{Hiromitsu Hori and Shoji Yamamoto}
\address{Division of Physics, Hokkaido University,
         Sapporo 060-0810, Japan}
\date{Received 4 March 2004}
\maketitle
\begin{abstract}
Nuclear spin-lattice relaxation in one-dimensional Heisenberg ferrimagnets
is studied by means of a modified spin-wave theory.
We consider the
second-order process, where a nuclear spin flip induces {\it virtual spin
waves} which are then scattered thermally via the four-magnon exchange
interaction, as well as the first-order process, where a nuclear spin
directly interacts with spin waves via the hyperfine interaction.
We point out {\it a possibility of the three-magnon relaxation process
predominating over the Raman one} and suggest model experiments.
\end{abstract}
\narrowtext

   Design of molecule-based ferromagnets is a challenging topic in
materials science \cite{K95} and numerous quasi-one-dimensional
ferrimagnets have been synthesized in this context.
Bimetallic chain compounds \cite{G7373,P138} are early examples,
one of which has indeed accomplished the three-dimensional ferromagnetic
order.
Another approach \cite{C1756} consists of bringing into interaction metal
ions and stable organic radicals.
Genuine organic ferrimagnets \cite{S1961,H7921} were also synthesized.
The tetrameric bond-alternating chain compound
Cu(C$_5$H$_4$NCl)$_2$(N$_3$)$_2$ \cite{E4466} and the trimeric
intertwining double-chain material Ca$_3$Cu$_3$(PO$_4$)$_4$ \cite{D83}
are distinct ferrimagnets of topological origin.

   Theoretical understanding of one-dimensional ferrimagnets has also
significantly been developed
\cite{V5144,D10992,P8894,Y14008,Y1024,O8067}, but very little is known
about their dynamic properties yet \cite{Y3711,H054409}.
Although the one-dimensional spin dynamics is a modern topic of great
interest \cite{V136}, systematic calculation of the nuclear magnetic
relaxation is still absent.
In the late sixties, Pincus and Beeman \cite{P398} formulated the nuclear
spin-lattice relaxation rate for Heisengerg ferromagnets and
antiferromagnets by means of the spin-wave theory and pointed out a
significant contribution of the three-magnon scattering to the relaxation
rate.
However, their theory is not effective in one dimension but is valid at
the onset of the three-dimensional long-range order.
In such circumstances, modifying the conventional spin-wave theory, we
make our first attempt at systematically describing the one-dimensional
nuclear spin dynamics on the basis of the spin-wave picture.
We predict that {\it the three-magnon relaxation process should
predominate over the Raman one at high temperatures and weak fields in
one-dimensional Heisenberg ferrimagnets}.

\begin{figure}
\centerline
{\mbox{\psfig{figure=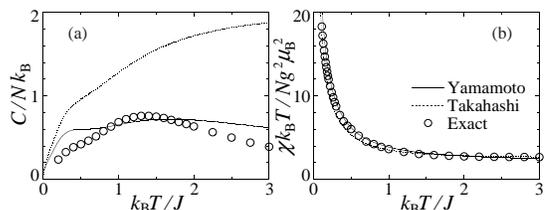,width=72mm,angle=0}}}
\caption{Modified spin-wave calculations of the specific heat and the
         magnetic susceptibility as functions of temperature for the
         spin-$(\frac{5}{2},\frac{1}{2})$ ferrimagnetic Heisenberg chain.
         The original (Takahashi) and our new (Yamamoto) schemes are
         compared with numerical findings (Exact) [10].}
\label{F:MSWdemo}
\end{figure}

   First of all our scheme \cite{Y14008,Y157603} of modifying the
spin-wave theory is distinct from the original idea proposed by Takahashi
\cite{T2494} and Hirsch {\it et al.} \cite{H4769}.
In the original way of suppressing the divergence of the sublattice
magnetizations, a Lagrange multiplier is first introduced and then the
effective Hamiltonian is diagonalized subject to zero staggered
magnetization.
The thus-obtained energy spectrum depends on temperature and fails to
describe the Schottky peak of the specific heat \cite{Y064426}.
In order to obtain better thermodynamics, we first diagonalize the
Hamiltonian, keeping the dispersion relations free from temperature, and
then introduce a Lagrange multiplier so as to minimize the free energy.
This scheme works well for ferrimagnets, as demonstrated in
Fig. \ref{F:MSWdemo}.
Both the approaches well reproduce the magnetic susceptibilty, but
our new shceme is much better than the original one at describing the
specific heat.
The nonvanishing specific heat at high temperatures, that is, the
endlessly increasing energy with temperature, is a fatal weak point of the
original scheme.

   We consider Heisenberg ferrimagnetic chains of alternating spins $S$
and $s$, as described by the Hamiltonian
\begin{equation}
   {\cal H}
      =\sum_{n=1}^N
       \Bigl[
        J\mbox{\boldmath$S$}_{n}\cdot
         \bigl(
          \mbox{\boldmath$s$}_{n-1}+\mbox{\boldmath$s$}_{n}
         \bigr)
       -g\mu_{\rm B}H\bigl(S_n^z+s_n^z\bigr)
       \Bigr]\,.
   \label{E:H}
\end{equation}
\vspace*{-5mm}
\vspace*{-7mm}
\begin{figure}
\centerline
{\mbox{\psfig{figure=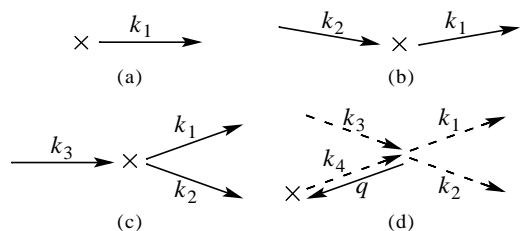,width=68mm,angle=0}}}
\caption{Illustration of the elementary nuclear spin-lattice relaxation
         processes.
         Solid arrows designate ferromagnetic or antiferromagnetic spin
         waves inducing a nuclear spin flip ($\times$).
         Broken arrows denote the four-magnon exchange interaction.
         (a) The direct process.
         (b) The first-order Raman process.
         (c) The first-order three-magnon process.
         (d) The second-order three-magnon process, where
             $q=-k_4\equiv k_3-k_2-k_1$.}
\label{F:diagram}
\end{figure}
\noindent
Introducing the Fourier-transformed Holstein-Primakoff magnon operators
$a_k$ and $b_k$ for each sublattice, we expand the Hamiltonian with
respect to $1/S$ as
${\cal H}={\cal H}_2+{\cal H}_1+{\cal H}_0+O(S^i)$, where we assume that
$O(S)=O(s)$ and ${\cal H}_i$ contains the $O(S^i)$ terms.
The $O(S^{-1})$ contributions are neglected in the following.
The hyperfine interaction is assumed to be of the dipolar type as
\begin{eqnarray}
   &&
   {\cal H}_{\rm hf}
   =g\mu_{\rm B}\hbar\gamma_{\rm N}I^+
   \nonumber\\
   &&\qquad\times
   \sum_{n=1}^N
    \Bigl[
     \frac{1}{2}\bigl(A_n^-S_n^-+B_n^-s_n^-\bigr)
    +A_n^zS_n^z+B_n^zs_n^z
    \Bigr],
\end{eqnarray}
where $A_n^\sigma$ and $B_n^\sigma$ are the hyperfine coupling tensors
between the nuclear and electronic spins.

   Since ${\cal H}_0$ and ${\cal H}_{\rm hf}$ are both much smaller than
${\cal H}_1$, we may regard ${\cal H}_0+{\cal H}_{\rm hf}\equiv{\cal V}$
as perturbative interactions to the linear spin-wave system
\begin{eqnarray}
   &&
   {\cal H}_2+{\cal H}_1
   =-2SsJN-(S+s)JN-g\mu_{\rm B}H(S-s)N
   \nonumber\\
   &&\qquad
   +J\sum_k\omega_k
   +J\sum_k
     \bigl(
      \omega_k^-\alpha_k^\dagger\alpha_k
     +\omega_k^+\beta_k^\dagger\beta_k
     \bigr),
\end{eqnarray}
where
$\alpha_k^\dagger
 \equiv a_k^\dagger{\rm cosh}\theta_k+b_k{\rm sinh}\theta_k$ and
$\beta_k^\dagger
 \equiv a_k{\rm sinh}\theta_k+b_k^\dagger{\rm cosh}\theta_k$,
provided ${\rm tanh}2\theta_k=2\sqrt{Ss}\cos(k/2)/(S+s)$, create the
ferromagnetic and antiferromagnetic spin waves of momentum $k$,
respectively, whose energies are given by
$\omega_k^\sigma=\omega_k+\sigma(S-s)-\sigma g\mu_{\rm B}H$ with
$\omega_k\equiv[(S-s)^2+4Ss\sin^2(k/2)]^{1/2}$.
We calculate the thermal distribution functions,
$\langle\alpha_k^\dagger\alpha_k\rangle\equiv\bar{n}_k^-$ and
$\langle\beta_k^\dagger\beta_k\rangle\equiv\bar{n}_k^+$,
by enforcing the zero staggered-magnetization constraint \cite{Y14008}.
If we consider up to the second-order perturbation with respect to
${\cal V}$, the probability of a nuclear spin being scattered from the
state of $I^z=m$ to that of $I^z=m+1$ is expressed as
\begin{equation}
   W=\frac{2\pi}{\hbar}\sum_f
     \Biggl|\Bigl\langle f\Bigl|
      {\cal V}+\sum_{m(\neq i)}
      \frac{{\cal V}|m\rangle\langle m|{\cal V}}{E_i-E_m}
     \Bigr|i\Bigr\rangle\Biggr|^2
     \delta(E_i-E_f),
   \label{E:W}
\end{equation}
where $i$ and $f$ designate the initial and final states of the
electronic-nuclear spin system described by the unperturbed Hamiltonian.
The nuclear spin-lattice relaxation time is then given by
$T_1=(I-m)(I+m+1)/2W$.
Equation (\ref{E:W}) contains various elementary relaxation processes,
which are illustrated in Fig. \ref{F:diagram}.
Due to the electronic-nuclear energy conservation, the direct process,
involving a single spin wave, is rarely of significance.
Within the first-order mechanism, where a nuclear spin directly interacts
with spin waves via the hyperfine interaction, the three-magnon relaxation
rate is much smaller than the Raman one \cite{P398}.
However, the first-order relaxation rates are generally enhanced through
the second-order mechanism, where a nuclear spin flip induces {\it virtual
spin waves} which are then scattered thermally via the four-magnon
exchange interaction.
We consider the leading second-order relaxation, that is, the
exchange-scattering-induced three-magnon process, as well as the
first-order relaxation.
The second-order relaxation of the Raman type originates in two virtual
spin waves within the present model and is therefore negligible in our
argument.

   Because of the significant difference between the electronic and
nuclear energy scales, $\hbar\omega_{\rm N}\ll J\omega_k^-<J\omega_k^+$,
the intraband spin-wave scatterings determine the Raman relaxation rate
$1/T_1^{(2)}$, whereas both the intraband and interband spin-wave
scatterings are relevant to the three-magnon relaxation rate
$1/T_1^{(3)}$.
Assuming the Fourier components of the coupling constants to have little
momentum dependence as
$\sum_n{\rm e}^{{\rm i}kn}A_n^\sigma\equiv A_k^\sigma\simeq A^\sigma$ and
$\sum_n{\rm e}^{{\rm i}kn}B_n^\sigma\equiv B_k^\sigma\simeq B^\sigma$,
we obtain
\begin{eqnarray}
   &&
   \frac{1}{T_1^{(2)}}
   \simeq\frac{2(g\mu_{\rm B}\hbar\gamma_{\rm N}B^z)^2}{\hbar JN}
    \sum_{k_1}\sum_{\sigma=\pm}\sum_{\tau=\pm}
     X^\sigma(\tau k_2^\sigma,k_1)^2
   \nonumber\\
   &&\qquad\times
     \bar{n}_{k_1}^\sigma(\bar{n}_{k_2^\sigma}^\sigma+1)
     \left|
      \frac{{\rm d}\omega_k^\sigma}{{\rm d}k}
     \right|_{k=k_2^\sigma}^{-1},
   \label{E:T1(2)}
   \\
   &&
   \frac{1}{T_1^{(3)}}
   \simeq\frac{(g\mu_{\rm B}\hbar\gamma_{\rm N}B^-)^2}{16\hbar SJN^2}
    \sum_{k_1,k_2}\sum_{\sigma=\pm}\sum_{\tau=\pm}
     2^{\sigma(\sigma-1)/2}
   \nonumber\\
   &&\qquad\times
     Y^\sigma(\tau k_3^\sigma,k_2,k_1)^2
     \bar{n}_{k_1}^\sigma\bar{n}_{k_2}^-(\bar{n}_{k_3^\sigma}^\sigma+1)
     \left|
      \frac{{\rm d}\omega_k^\sigma}{{\rm d}k}
     \right|_{k=k_3^\sigma}^{-1},
   \label{E:T1(3)}
\end{eqnarray}
where $k_2^\sigma$ and $k_3^\sigma$ are given by
$\omega_{k_1}^\sigma-\omega_{k_2^\sigma}^\sigma-\hbar\omega_{\rm N}/J=0$
and
$\omega_{k_1}^\sigma+\omega_{k_2}^-
-\omega_{k_3^\sigma}^\sigma-\hbar\omega_{\rm N}/J=0$,
respectively, and
\widetext
\begin{eqnarray}
   &&
   X^-(k_1,k_2)
   =\frac{A^z}{B^z}{\rm ch}\theta_{k_1}{\rm ch}\theta_{k_2}
   -               {\rm sh}\theta_{k_1}{\rm sh}\theta_{k_2},\ \ 
   X^+(k_1,k_2)
   =\frac{A^z}{B^z}{\rm sh}\theta_{k_1}{\rm sh}\theta_{k_2}
   -               {\rm ch}\theta_{k_1}{\rm ch}\theta_{k_2},
   \\
   &&
   Y^-(k_1,k_2,k_3)
   =\frac{A^-}{B^-}
    {\rm ch}\theta_{k_1}{\rm ch}\theta_{k_2}{\rm ch}\theta_{k_3}
   -\sqrt{\frac{S}{s}}
    {\rm sh}\theta_{k_1}{\rm sh}\theta_{k_2}{\rm sh}\theta_{k_3}
   -\frac{2SV_1(k_1,k_2,k_3,k_3+k_2-k_1)}
         {J\omega_{k_3+k_2-k_1}^--\hbar\omega_{\rm N}}
   \nonumber\\
   &&\quad\times
    \Bigl(
     \frac{A^-}{B^-}{\rm ch}\theta_{k_3+k_2-k_1}
    -\sqrt{\frac{s}{S}}{\rm sh}\theta_{k_3+k_2-k_1}
    \Bigr)
   -\frac{ SV_2(k_1,k_2,k_3,k_3+k_2-k_1)}
         {J\omega_{k_3+k_2-k_1}^++\hbar\omega_{\rm N}}
    \Bigl(
     \frac{A^-}{B^-}{\rm sh}\theta_{k_3+k_2-k_1}
    -\sqrt{\frac{s}{S}}{\rm ch}\theta_{k_3+k_2-k_1}
    \Bigr),
   \nonumber\\
   &&
   Y^+(k_1,k_2,k_3)
   =2\Bigl(
     \frac{A^-}{B^-}
     {\rm sh}\theta_{k_1}{\rm ch}\theta_{k_2}{\rm sh}\theta_{k_3}
    -\sqrt{\frac{S}{s}}
     {\rm ch}\theta_{k_1}{\rm sh}\theta_{k_2}{\rm ch}\theta_{k_3}
     \Bigr)
   -\frac{ SV_3(k_3+k_2-k_1,k_3,k_2,k_1)}
         {J\omega_{k_3+k_2-k_1}^--\hbar\omega_{\rm N}}
   \nonumber\\
   &&\quad\times
    \Bigl(
     \frac{A^-}{B^-}{\rm ch}\theta_{k_3+k_2-k_1}
    -\sqrt{\frac{s}{S}}{\rm sh}\theta_{k_3+k_2-k_1}
    \Bigr)
   -\frac{2SV_4(k_3,k_3+k_2-k_1,k_1,k_2)}
         {J\omega_{k_3+k_2-k_1}^++\hbar\omega_{\rm N}}
    \Bigl(
     \frac{A^-}{B^-}{\rm sh}\theta_{k_3+k_2-k_1}
    -\sqrt{\frac{s}{S}}{\rm ch}\theta_{k_3+k_2-k_1}
    \Bigr).
\end{eqnarray}
Here, ${\rm ch}\theta_k\equiv{\rm cosh}\theta_k$,
${\rm sh}\theta_k\equiv{\rm sinh}\theta_k$, and the four-magnon exchange
interaction $V_i$ is given as
\begin{eqnarray}
   &&
   \frac{V_1(k_1,k_2,k_3,k_4)}{J}=
   \Bigl(\cos\!\frac{k_4-k_2}{2}+\cos\!\frac{k_3-k_1}{2}\Bigr)
   \bigl(
    {\rm ch}\theta_{k_1}{\rm sh}\theta_{k_2}
    {\rm ch}\theta_{k_3}{\rm sh}\theta_{k_4}
   +{\rm sh}\theta_{k_1}{\rm ch}\theta_{k_2}
    {\rm sh}\theta_{k_3}{\rm ch}\theta_{k_4}
   \bigr)
   \nonumber\\
   &&\quad
  +\Bigl(\cos\!\frac{k_4-k_3}{2}+\cos\!\frac{k_2-k_1}{2}\Bigr)
   \bigl(
    {\rm ch}\theta_{k_1}{\rm ch}\theta_{k_2}
    {\rm sh}\theta_{k_3}{\rm sh}\theta_{k_4}
   +{\rm sh}\theta_{k_1}{\rm sh}\theta_{k_2}
    {\rm ch}\theta_{k_3}{\rm ch}\theta_{k_4}
   \bigr)
   \nonumber\\
   &&\quad
  -\sqrt{\frac{S}{s}}
   \Bigl[
   {\rm sh}\theta_{k_3}{\rm sh}\theta_{k_4}
   \Bigl(
    \cos\!\frac{k_1}{2}{\rm ch}\theta_{k_1}{\rm sh}\theta_{k_2}
   +\cos\!\frac{k_2}{2}{\rm sh}\theta_{k_1}{\rm ch}\theta_{k_2}
   \Bigr)
   +                 {\rm sh}\theta_{k_1}{\rm sh}\theta_{k_2}
   \Bigl(
    \cos\!\frac{k_3}{2}{\rm ch}\theta_{k_3}{\rm sh}\theta_{k_4}
   +\cos\!\frac{k_4}{2}{\rm sh}\theta_{k_3}{\rm ch}\theta_{k_4}
   \Bigr)
   \Bigr]
   \nonumber\\
   &&\quad
  -\sqrt{\frac{s}{S}}
   \Bigl[
   {\rm ch}\theta_{k_3}{\rm ch}\theta_{k_4}
   \Bigl(
    \cos\!\frac{k_1}{2}{\rm sh}\theta_{k_1}{\rm ch}\theta_{k_2}
   +\cos\!\frac{k_2}{2}{\rm ch}\theta_{k_1}{\rm sh}\theta_{k_2}
   \Bigr)
   +                 {\rm ch}\theta_{k_1}{\rm ch}\theta_{k_2}
   \Bigl(
    \cos\!\frac{k_3}{2}{\rm sh}\theta_{k_3}{\rm ch}\theta_{k_4}
   +\cos\!\frac{k_4}{2}{\rm ch}\theta_{k_3}{\rm sh}\theta_{k_4}
   \Bigr)
   \Bigr],
   \nonumber\\
   &&
   \frac{V_2(k_1,k_2,k_3,k_4)}{2J}=
   \Bigl(\cos\!\frac{k_4-k_2}{2}+\cos\!\frac{k_3-k_1}{2}\Bigr)
   \bigl(
    {\rm ch}\theta_{k_1}{\rm sh}\theta_{k_2}
    {\rm ch}\theta_{k_3}{\rm ch}\theta_{k_4}
   +{\rm sh}\theta_{k_1}{\rm ch}\theta_{k_2}
    {\rm sh}\theta_{k_3}{\rm sh}\theta_{k_4}
   \bigr)
   \nonumber\\
   &&\quad
  +\Bigl(\cos\!\frac{k_4-k_3}{2}+\cos\!\frac{k_2-k_1}{2}\Bigr)
   \bigl(
    {\rm ch}\theta_{k_1}{\rm ch}\theta_{k_2}
    {\rm sh}\theta_{k_3}{\rm ch}\theta_{k_4}
   +{\rm sh}\theta_{k_1}{\rm sh}\theta_{k_2}
    {\rm ch}\theta_{k_3}{\rm sh}\theta_{k_4}
   \bigr)
   \nonumber\\
   &&\quad
  -\sqrt{\frac{S}{s}}
   \Bigl[
   {\rm sh}\theta_{k_3}{\rm ch}\theta_{k_4}
   \Bigl(
    \cos\!\frac{k_1}{2}{\rm ch}\theta_{k_1}{\rm sh}\theta_{k_2}
   +\cos\!\frac{k_2}{2}{\rm sh}\theta_{k_1}{\rm ch}\theta_{k_2}
   \Bigr)
   +                 {\rm sh}\theta_{k_1}{\rm sh}\theta_{k_2}
   \Bigl(
    \cos\!\frac{k_3}{2}{\rm ch}\theta_{k_3}{\rm ch}\theta_{k_4}
   +\cos\!\frac{k_4}{2}{\rm sh}\theta_{k_3}{\rm sh}\theta_{k_4}
   \Bigr)
   \Bigr]
   \nonumber\\
   &&\quad
  -\sqrt{\frac{s}{S}}
   \Bigl[
   {\rm ch}\theta_{k_3}{\rm sh}\theta_{k_4}
   \Bigl(
    \cos\!\frac{k_1}{2}{\rm sh}\theta_{k_1}{\rm ch}\theta_{k_2}
   +\cos\!\frac{k_2}{2}{\rm ch}\theta_{k_1}{\rm sh}\theta_{k_2}
   \Bigr)
   +                 {\rm ch}\theta_{k_1}{\rm ch}\theta_{k_2}
   \Bigl(
    \cos\!\frac{k_3}{2}{\rm sh}\theta_{k_3}{\rm sh}\theta_{k_4}
   +\cos\!\frac{k_4}{2}{\rm ch}\theta_{k_3}{\rm ch}\theta_{k_4}
   \Bigr)
   \Bigr],
   \nonumber\\
   &&
   \frac{V_3(k_1,k_2,k_3,k_4)}{4J}=
   \Bigl(\cos\!\frac{k_4-k_2}{2}+\cos\!\frac{k_3-k_1}{2}\Bigr)
   \bigl(
    {\rm ch}\theta_{k_1}{\rm ch}\theta_{k_2}
    {\rm ch}\theta_{k_3}{\rm ch}\theta_{k_4}
   +{\rm sh}\theta_{k_1}{\rm sh}\theta_{k_2}
    {\rm sh}\theta_{k_3}{\rm sh}\theta_{k_4}
   \bigr)
   \nonumber\\
   &&\quad
  +\Bigl(\cos\!\frac{k_4-k_3}{2}+\cos\!\frac{k_2-k_1}{2}\Bigr)
   \bigl(
    {\rm ch}\theta_{k_1}{\rm sh}\theta_{k_2}
    {\rm sh}\theta_{k_3}{\rm ch}\theta_{k_4}
   +{\rm sh}\theta_{k_1}{\rm ch}\theta_{k_2}
    {\rm ch}\theta_{k_3}{\rm sh}\theta_{k_4}
   \bigr)
   \nonumber\\
   &&\quad
  -\sqrt{\frac{S}{s}}
   \Bigl[
   {\rm sh}\theta_{k_3}{\rm ch}\theta_{k_4}
   \Bigl(
    \cos\!\frac{k_1}{2}{\rm ch}\theta_{k_1}{\rm ch}\theta_{k_2}
   +\cos\!\frac{k_2}{2}{\rm sh}\theta_{k_1}{\rm sh}\theta_{k_2}
   \Bigr)
   +                 {\rm sh}\theta_{k_1}{\rm ch}\theta_{k_2}
   \Bigl(
    \cos\!\frac{k_3}{2}{\rm ch}\theta_{k_3}{\rm ch}\theta_{k_4}
   +\cos\!\frac{k_4}{2}{\rm sh}\theta_{k_3}{\rm sh}\theta_{k_4}
   \Bigr)
   \Bigr]
   \nonumber\\
   &&\quad
  -\sqrt{\frac{s}{S}}
   \Bigl[
   {\rm ch}\theta_{k_3}{\rm sh}\theta_{k_4}
   \Bigl(
    \cos\!\frac{k_1}{2}{\rm sh}\theta_{k_1}{\rm sh}\theta_{k_2}
   +\cos\!\frac{k_2}{2}{\rm ch}\theta_{k_1}{\rm ch}\theta_{k_2}
   \Bigr)
   +                 {\rm ch}\theta_{k_1}{\rm sh}\theta_{k_2}
   \Bigl(
    \cos\!\frac{k_3}{2}{\rm sh}\theta_{k_3}{\rm sh}\theta_{k_4}
   +\cos\!\frac{k_4}{2}{\rm ch}\theta_{k_3}{\rm ch}\theta_{k_4}
   \Bigr)
   \Bigr],
   \nonumber\\
   &&
   \frac{V_4(k_1,k_2,k_3,k_4)}{2J}=
   \Bigl(\cos\!\frac{k_4-k_2}{2}+\cos\!\frac{k_3-k_1}{2}\Bigr)
   \bigl(
    {\rm ch}\theta_{k_1}{\rm sh}\theta_{k_2}
    {\rm ch}\theta_{k_3}{\rm ch}\theta_{k_4}
   +{\rm sh}\theta_{k_1}{\rm ch}\theta_{k_2}
    {\rm sh}\theta_{k_3}{\rm sh}\theta_{k_4}
   \bigr)
   \nonumber\\
   &&\quad
  +\Bigl(\cos\!\frac{k_4-k_3}{2}+\cos\!\frac{k_2-k_1}{2}\Bigr)
   \bigl(
    {\rm ch}\theta_{k_1}{\rm ch}\theta_{k_2}
    {\rm sh}\theta_{k_3}{\rm ch}\theta_{k_4}
   +{\rm sh}\theta_{k_1}{\rm sh}\theta_{k_2}
    {\rm ch}\theta_{k_3}{\rm sh}\theta_{k_4}
   \bigr)
   \nonumber\\
   &&\quad
  -\sqrt{\frac{S}{s}}
   \Bigl[
   {\rm ch}\theta_{k_3}{\rm sh}\theta_{k_4}
   \Bigl(
    \cos\!\frac{k_1}{2}{\rm sh}\theta_{k_1}{\rm ch}\theta_{k_2}
   +\cos\!\frac{k_2}{2}{\rm ch}\theta_{k_1}{\rm sh}\theta_{k_2}
   \Bigr)
   +                 {\rm ch}\theta_{k_1}{\rm ch}\theta_{k_2}
   \Bigl(
    \cos\!\frac{k_3}{2}{\rm sh}\theta_{k_3}{\rm sh}\theta_{k_4}
   +\cos\!\frac{k_4}{2}{\rm ch}\theta_{k_3}{\rm ch}\theta_{k_4}
   \Bigr)
   \Bigr]
   \nonumber\\
   &&\quad
  -\sqrt{\frac{s}{S}}
   \Bigl[
   {\rm sh}\theta_{k_3}{\rm ch}\theta_{k_4}
   \Bigl(
    \cos\!\frac{k_1}{2}{\rm ch}\theta_{k_1}{\rm sh}\theta_{k_2}
   +\cos\!\frac{k_2}{2}{\rm sh}\theta_{k_1}{\rm ch}\theta_{k_2}
   \Bigr)
   +                 {\rm sh}\theta_{k_1}{\rm sh}\theta_{k_2}
   \Bigl(
    \cos\!\frac{k_3}{2}{\rm ch}\theta_{k_3}{\rm ch}\theta_{k_4}
   +\cos\!\frac{k_4}{2}{\rm sh}\theta_{k_3}{\rm sh}\theta_{k_4}
   \Bigr)
   \Bigr].
\end{eqnarray}
\narrowtext

\begin{figure}
\centerline
{\mbox{\psfig{figure=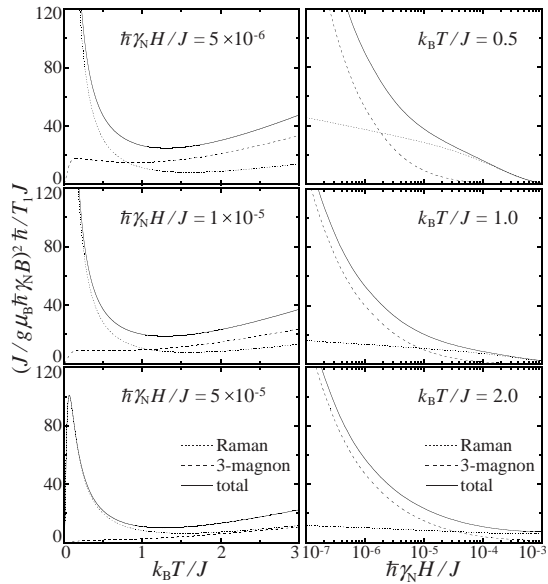,width=72mm,angle=0}}}
\caption{Modified spin-wave calculations of the temperature (the left
         three) and field (the right three) dependences of the nuclear
         spin-lattice relaxation rate at $A/B=0.1$, where $1/T_1^{(2)}$
         and $1/T_1^{(3)}$ are plotted by dotted and broken lines,
         respectively, while $1/T_1^{(2)}+1/T_1^{(3)}\equiv 1/T_1$,
         which is observable, by solid lines.}
\label{F:T1}
\end{figure}

   We calculate the case of $(S,s)=(\frac{5}{2},\frac{1}{2})$, which is
relevant to several major materials \cite{G7373,P138,C1756}, assuming
that $A^-/2\simeq A^z\equiv A$ and $B^-/2\simeq B^z\equiv B$.
Figure \ref{F:T1} shows $1/T_1$ as a function of temperature and an
applied magnetic field.
All in all, the exchange-scattering-enhanced three-magnon relaxation rate
$1/T_1^{(3)}$ grows with increasing temperature and decreasing field and
ends in the leading contribution to $1/T_1$.
The temperature dependence of $1/T_1$ is determined by $\bar{n}_k^-$
unless temperature is sufficiently high.
As temperature increases, $\bar{n}_k^-$ is decreasing at
$k\simeq 0$ but otherwise increasing \cite{Y2324}.
$|{\rm d}\omega_k^-/{\rm d}k|_{k=k_2^-}^{-1}$ makes the $k_1\simeq 0$
components predominant in the momentum summation (\ref{E:T1(2)}),
while such predominance arising from
$|{\rm d}\omega_k^-/{\rm d}k|_{k=k_3^-}^{-1}$ is suppressed in the double
summation (\ref{E:T1(3)}).
$1/T_1^{(2)}$ and $1/T_1^{(3)}$ are hence decreasing and increasing,
respectively, with increasing temperature and therefore, the
Raman-to-three-magnon crossover may be detected at low fields.
   
\begin{figure}
\centerline
{\mbox{\psfig{figure=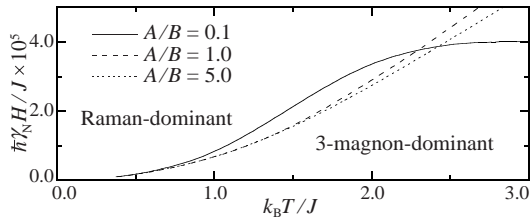,width=70mm,angle=0}}}
\caption{The crossover point, where $1/T_1^{(2)}=1/T_1^{(3)}$, as a
         function of temperature and an applied field.}
\label{F:PhD}
\end{figure}

   The field dependence of $1/T_1$ is also useful in detecting the
crossover.
At moderately low temperatures and weak fields,
$\hbar\omega_{\rm N}\ll k_{\rm B}T\ll J$, we may evaluate Eq.
(\ref{E:T1(2)}) as
\begin{equation}
   \frac{1}{T_1^{(2)}}
   \simeq
   \frac{2(g\mu_{\rm B}\hbar\gamma_{\rm N})^2(AS-Bs)^2}{\pi\hbar Ss(S-s)J}
   {\rm e}^{-g\mu_{\rm B}H/k_{\rm B}T}
   K_0\Bigl(\frac{\hbar\omega_{\rm N}}{2k_{\rm B}T}\Bigr),
\end{equation}
where $K_0$ is the modified Bessel function of the second kind and
behaves as
$K_0(\hbar\omega_{\rm N}/2k_{\rm B}T)
 \simeq 0.80908-{\rm ln}(\hbar\omega_{\rm N}/k_{\rm B}T)$
for $\hbar\omega_{\rm N}\ll k_{\rm B}T$.
Considering the relation $\omega_{\rm N}=\gamma_{\rm N}H$, we find that
the field dependence of $1/T_1^{(2)}$ is logarithmic at weak fields
and turns exponential with increasing field.
Since $1/T_1^{(3)}$ exhibits much stronger initial field dependence, the
three-magnon relaxation can be detected at weak fields.
   We plot {\it the crossover points} in Fig. \ref{F:PhD}.
The three-magnon relaxation process generally predominates over the
Raman one at high temperatures and$\,$ weak$\,$ fields.$\,$
Since$\,$ the$\,$ ferrimagnetic$\,$ nuclear$\,$ spin-
\vspace*{-4mm}
\noindent
lattice relaxation is very sensitive
to another adjustable parameter $A/B$, that is, the location of the probe
nuclei, any $T_1$ measurement should be accompanied by detailed structural
analyses.
At the special location of $A/B\sim(d_s/d_S)^3\simeq s/S$, where $d_S$
($d_s$) is the distance between the nuclear and larger (smaller)
electronic spins, the leading ferromagnetic spin waves are almost
invisible to the nuclear spins \cite{H054409,Y2324} and therefore $1/T_1$
is too small to be measured.
The three-magnon-dominant relaxation rate is detectable for
$A/B=0.1$ and $A/B=5.0$, for exmaple,
which are relevant to $^1$H nuclei in
MnCu(pbaOH)(H$_2$O)$_3$ with $J/k_{\rm B}\simeq 34\,\mbox{K}$ \cite{P138}
and $^{19}$F nuclei in
Mn(hfac)$_2$NITiPr with $J/k_{\rm B}\simeq 460\,\mbox{K}$ \cite{C1756},
respectively.
When we measure MnCu(pbaOH)(H$_2$O)$_3$ and Mn(hfac)$_2$NITiPr at
$100\,\mbox{K}$ and room temperature, respectively, the three-magnon
relaxation is possibly dominant for $H\alt 0.6\,\mbox{T}$ in both
compounds.

   There exist pioneering $T_1$ measurements on the layered ferromagnet
CrCl$_3$ \cite{N354} and the coupled-chain antiferromagnet
CsMnCl$_3\cdot$2H$_2$O \cite{N5325}, which give evidence of the relevant
three-magnon relaxation process in accordance with the Pincus-Beeman
formulation \cite{P398}.
However, they are both low-temperature measurements under the existing
three-dimensional long-range order.
A similarly-motivated experiment on the one-dimensional easy-plane
ferromagnet CsNiF$_3$
\cite{G6347} was performed at higher temperatures but explained in terms of
solitons, possibly because of the lack of the spin-wave formulation in one
dimension.
Here is a modern theoretical tool$-$the modified spin-wave theory based on
the new scheme \cite{Y14008}.
Let us observe the three-magnon nuclear-spin scattering in the
one-dimensional quantum relaxation.

   The authors are grateful to Professor T. Goto for valuable comments.
This work was supported by the Ministry of Education, Culture, Sports,
Science and Technology of Japan and the Nissan Science Foundation.

\widetext
\end{document}